\documentclass[aps,prb,twocolumn,superscriptaddress,showpacs,amsmath,amssymb,footinbib,longbibliography]{revtex4-2}
\usepackage[english]{babel}
\usepackage{amssymb,amsbsy,amsmath}
\usepackage{bbm}
\usepackage{graphicx}
\usepackage{dcolumn}
\usepackage{bm}
\usepackage{subfigure}
\usepackage{color}
\usepackage{textcomp}
\usepackage[colorlinks,bookmarks=false,citecolor=blue,linkcolor=red,urlcolor=blue]{hyperref}

\newcommand{\figref}[1]{Fig.~\protect\ref{#1}}

\begin{document}
\title{The magnetization process of classical  Heisenberg magnets with
non-coplanar cuboc ground states}

\author{Johannes Richter}
\affiliation{Institut f\"ur Physik, Universit\"at Magdeburg, P.O. Box 4120, D-39016 Magdeburg, Germany}
\author{Heinz-J\"{u}rgen Schmidt}
\email{hschmidt@uos.de}
\affiliation{Fachbereich Mathematik, Informatik, Physik, Universit\"{a}t Osnabr\"{u}ck, 49069 Osnabr\"{u}ck, Germany.}
\author{J\"{u}rgen Schnack}
\affiliation{Fakult\"at f\"ur Physik, Universit\"{a}t Bielefeld, P.O. Box 100131, 33501 Bielefeld, Germany.}

\date{\today}

\begin{abstract}

We consider a classical Heisenberg model on
the kagom\'{e} and the square kagom\'{e} lattice, where at zero magnetic field
non-coplanar cuboctahedral ground states with twelve sublattices
exist if suitable exchange couplings are introduced between the other neighbors.
Such 'cuboc ground states' are remarkable because they allow for chiral ordering.
For these models, we discuss the magnetization process
in an applied magnetic field $H$ by both numerical and analytical methods.
We find some universal properties that are present in all models.
The magnetization curve $M(H)$ usually contains only non-linear components
and there is at least one magnetic field driven phase transition.
Details of the $M(H)$ curve such as the number and characteristics
(continuous or discontinuous) of the phase transitions depend on the lattice
and the details of the exchange between the further neighbors.
Typical features of these magnetization processes can already
be derived for a paradigmatic 12 spin model that we define in this work.

\end{abstract}

\pacs{75.10.Jm,75.50.Xx,75.40.Mg} \keywords{Kagome, Classical Heisenberg Model, Frustration, Magnetization}

\maketitle

\section{Introduction}
\label{sec-1}

Heisenberg antiferromagnets  on two-dimensional lattices of corner-sharing
triangles
are the prototypes of highly frustrated magnets.
The celebrated kagom\'{e} antiferromagnet (KHAF) is the most prominent example.
Already in the classical limit the model exhibits unconventional
properties. The corner-sharing triangular geometry leads to a highly
degenerate
ground-state manifold including coplanar as well as non-coplanar spin
configurations \cite{Chalker1992,Harris1992,Huse:1992,Reimers1993}.
Quantum fluctuations select the coplanar $\sqrt{3} \times \sqrt{3}$ state
\cite{Sachdev_1992,Chubukov:92,Henley:1995,Goetze2011}.
The investigation of the low-temperature physics of the classical model is also
highly non-trivial since  the free energy exhibits many different local
minima with entropic barriers between them
\cite{Zhitomirsky2008,Moessner2013_class_kagome}.

Another spin system  on a lattice with corner-sharing triangles is
the square-kagom\'{e} Heisenberg antiferromagnet (SKHAF).
The model was introduced about 20 years ago
\cite{Sidd2001,schnack2001independent,tomczak2003specific,richter2004-spin-peierls,richter2009squago,Sakai2013,
Rousochatzakis2013,derzhko2014square,Rousochatzakis2015,Sakai2015,Derzhko2015review}.
Over the last 5
years the  SKHAF has attracted more and more attention on the
theoretical
\cite{Hasegawa2018,Morita2018,Lugan2019,McClarty2020,PhysRevB.102.241115,Iqbal2021,schmoll2022tensor,
squago_j1j2_2023,squago_clas_2023,decorated_squago_2023}
and experimental \cite{FMM:NC20,YSK:IC21,Vasiliev2022,Vasiliev2023,Murtazoev-2023} side.
The  SKHAF shares many properties with the
KHAF, such as a highly
degenerate ground-state manifold including coplanar $\sqrt{3} \times \sqrt{3}$
and $q=0$ states \cite{richter2009squago,Iqbal2021,squago_clas_2023}, the
absence of magnetic order in the quantum $s=1/2$ model
\cite{richter2009squago,Sakai2013,Rousochatzakis2013,Lugan2019,Iqbal2021,schmoll2022tensor},
and flat-band phenomena emerging in applied magnetic field
\cite{Derzhko2006universal,Derzhko2015review}.
Moreover,
the SKHAF and the KHAF exhibit similar thermodynamic properties for the quantum $s=1/2$
\cite{Xi-Chen2018,kago42,Hotta_2018,squago_TD_2022}
as well as for the classical model
\cite{Zhitomirsky2008,Moessner2013_class_kagome,squago_clas_2023}.

For both models  the massive degeneracy of the classical
ground-state manifold  can be lifted by additional exchange couplings such as
2nd-nearest neighbor and 3rd-nearest neighbor bonds, see, e.~g.,
\cite{Domenge2005_cuboc2,janson2008modified,Messio_2011_clas_GS,Valenti2015_kago_j1j2jd,Messio_kago_j1j3_PRB2020,squago_clas_2023}.
However, to find the classical ground state for such models sometimes appears
to be challenging, in particular if the Luttinger-Tisza method fails, see
e.g. \cite{Messio_kago_j1j3_PRB2020,squago_clas_2023,clas_GS_JPA_2022}.

Of particular interest is the quest for classical Heisenberg models with exclusively non-coplanar ground
states. In this case
finite-temperature phase transitions in two-dimensional Heisenberg models
can emerge, which do
not contradict the famous Mermin-Wagner theorem \cite{mermin1966absence}.
Prominent examples of non-coplanar classical ground
states on two-dimensional lattices are the so-called cuboc1, cuboc2 and
cuboc3
phases. These phases exhibit 12 non-coplanar sublattices
pointing
towards the 12 vertices of the cuboctahedron.
Such classical non-coplanar ground states are
candidates as parents for chiral spin liquids in corresponding quantum spin models
\cite{Domenge2005_cuboc2,PhysRevB.93.094437,PhysRevB.96.115115}.

The cuboc2 phase was found in
Ref.~\cite{Domenge2005_cuboc2} for the classical KHAF with ferromagnetic
nearest-neighbor
exchange $J_1$ and
and antiferromagnetic  2nd neighbor
exchange $J_2$.
The cuboc1 phase was first reported  in
Refs.~\cite{janson2008modified,janson2009intrinsic}
for the classical KHAF with antiferromagnetic nearest-neighbor
exchange $J_1$ and 3rd neighbor exchange $J_d$
along the diagonals of the
hexagons.
Later on, in
Ref.~\cite{Messio_2011_clas_GS} a systematic analysis of these phases was
given and the notations cuboc1 and cuboc2 were introduced.
All neighbouring pairs of spins form an angle of 120° (cuboc1) or 60° (cuboc2).

Very recently, cuboc phases  were found and analyzed also for the classical
SKHAF with further-neighbor exchange $J_+$ and
$J_{\rm x}$ along the
diagonals of the octagons of the square-kagom\'{e} lattice  \cite{squago_clas_2023}.
For this model, except  the cuboc1 phase already known from the  KHAF,
a new cuboc3 phase was detected,
featuring two different angles between neighbouring
spin pairs, 120° and 60°,  associated
with the two non-equivalent nearest-neighbor bonds
of the SKHAF.

In the present paper we focus on the ground-state magnetization process of classical
Heisenberg models having cuboc ground states in zero magnetic field.
The corresponding Hamiltonian  augmented with a Zeeman term is given by
\begin{equation}
\label{Ham}
{\cal H}
=
\hspace*{-2mm} \sum_{i<j}
J_{ij}{\vec s}_i\cdot{\vec s}_j \;
- H \;
\sum_{i} s^z_i \; , \;  |{\vec s}_i| =1
\ .
\end{equation}
For most classical Heisenberg antiferromagnets the ground-state magnetization $M=\sum_{i}
s^z_i$ increases linearly up to saturation $M_{\rm sat} = N$, where $N$ is
the number of spins.
Thus, for the KHAF and the SKHAF with only nearest-neighbor bonds $J_1$
the magnetization  is given by
$M/N=H/(6J_1)$, $H\le H_{\rm sat} = 6J_1$, see, e.g.. Ref.~\cite{zhito_PRL2002}.
On the contrary, the magnetization curves of the quantum KHAF and SKHAF
exhibit plateaus and jumps
\cite{Hida2001,schulenburg2002jump,honecker2004plateau,richter2009squago,Sakai2013}.
In general, the linearity of the classical magnetization curve may have different reasons.
In the case of the KHAF and the SKHAF with only nearest-neighbor bonds it
follows from the existence of coplanar ground states with vanishing total spin
for $H=0$.
As the magnetic field increases, the spin vectors are folded in the direction of the magnetic field axis,
much like the ribs of a closing umbrella.
The resulting $3$-dimensional configuration is therefore also called the ``umbrella construction".
In the cases of the KHAF and the SKHAF with additional bonds and non-coplanar zero-field cuboc ground states
the above-sketched umbrella construction is no longer possible.
Interesting enough, there exists an analogous, purely mathematical $4$-dimensional umbrella construction,
but this yields non-physical ground states \cite{Schmidt2017}.
It is thus plausible that the search for physical, i.~e., at most $3$-dimensional
ground states leads to different phases and nonlinear magnetization curves.

Over the last two decades, for the isotropic Heisenberg antiferromagnet
only a few examples of classical magnetization curves
with jumps were reported for finite systems with icosahedral symmetry
or fullerene molecules
\cite{PhysRevLett.69.176,SSS:PRL05,Kon:PRB05,Kon:PRB07,Konstantinidis2018,Konstantinidis2023,Konstantinidis2023a}.
However, quite recently in Refs.~\cite{Zhito_M_H_cuboc} and \cite{M_H_cuboc_I} unconventional
classical magnetization curves with jumps and kinks have been found for a
frustrated spinel, the $J_1$-$J_d$ KHAF as well as for the SKHAF with
further-neighbor bonds.
Significantly, in these cases there is always a $3$-dimensional ground state for $H=0$
with vanishing total spin, which, according to the above, makes sense.

In this study we discuss all classical spin lattices known to us in which a
cuboctahedral ground state is reported in the literature and compare their magnetization curves.
In particular, we consider four two-dimensional
classical spin-lattice models: the kagom\'{e} $J_1$-$J_2$ magnet
with $J_1<0$, $J_2>0$, the kagom\'{e} $J_1$-$J_d$ magnet with $J_1,J_d>0$
and the square-kagom\'{e} $J_1$-$J_+$-$J_{\rm x}$ magnet with   $J_1,J_+=J_{\rm x}>0$ and
with    $J_1<0$ and $J_+=J_{\rm x}>0$.
Note that in all these cases the lack of coplanar classical ground states,
i.e., the
existence of only non-coplaner ground states,  is related to
additional further-neighbor bonds.
The focus is here on the kagom\'{e} models since for the  kagom\'{e} $J_1$-$J_2$ magnet
so far no analysis of the classical magnetization curve is available.
We also  briefly
consider the kagom\'{e} $J_1$-$J_d$ magnet with $J_1,J_d>0$
and the square-kagom\'{e} $J_1$-$J_+$-$J_{\rm x}$ for comparison and add some
new data not presented in Refs.~\cite{Zhito_M_H_cuboc} and \cite{M_H_cuboc_I}.
Before we deal with these lattice models, we consider as an introductory and
characteristic example a classical Heisenberg model for a spin system with 12 spins,
in which the ground state is unique with a suitable arrangement of
three different exchange bonds and becomes a non-coplanar cuboc state.

For this finite system of $N=12$ spins we will
provide several analytical expressions for the magnetization curve.
We will also demonstrate that many features  of the magnetization of the
kagom\'{e} and square-kagom\'{e} spin lattices are already present in the simpler
12 spin model.
In addition, the small number of only $N=12$ spins allows quantum mechanical
calculations for spin quantum numbers up to $s=9/2$, which we use for comparison with the classical results.

\section{Methods}
\label{sec-2}

Let us briefly describe the used methods.
We use a variant of the iterative minimization method to get numerical ground-state data for the spin
configuration, the energy $E_0$ and the magnetization $M$. This method is described in more detail
in Ref.~\cite{M_H_cuboc_I}.

We also use a semi-analytical approach introduced in
\cite{squago_clas_2023,M_H_cuboc_I},
which we will describe here briefly for con\-venience.
We use numerical data as input to figure out groups  of spins approximately pointing into
the same direction. Based on the lattice symmetries we determine the symmetry group
of the resulting spin orientations,
which eventually allows to reduce the number of different spin orientations.
The energy $E$ is then written as a function of a few parameters $\alpha_i$,
i.e., $E={\mathcal H}(\alpha_1,\ldots,\alpha_n)$, and can be minimized
analytically or numerically.
The criterion for the success of the semi-analytical method is the lowering
of the resulting ground state energy compared to the corresponding numerical value.
This method enables the exact calculation of the boundaries between
individual ground state phases, and thus the construction of precise phase diagrams.

In special cases, typically for the last phase before saturation,
the number of free parameters for the ground states is so small that an
analytical calculation becomes possible.
Alternatively, a generalized Luttinger-Tisza method can be used
for this purpose \cite{clas_GS_JPA_2022}.
To illustrate our approach, in the next section we present both the semi-analytical
and the fully analytical solution for the 12 spin model in detail.

We often use so-called ``common origin plots" to visualize classical ground states.
In a common origin plot, the spin vectors of a ground state,
which belong to the different spin sites, are plotted in a single unit sphere.
Although this means that the exact information about the distribution of the spin vectors
on the lattice is lost, structures and symmetries
that characterize the ground state in more detail often become visible.

\section{The 12 spin model}
\label{sec-3}

The cuboctahedron is one of the 13 Archimedean solids and is created
by joining the centers of the 12 edges of the cube.
The set of rotations and reflections leaving the cuboctahedron invariant
generates a 'natural' 12-dimensional linear representation of the octahedral group ${\cal O}_h$
of order 48.

The Heisenberg model on the cuboctahedron
was studied for the quantum \cite{schmidt2005frustration,ScS:P09} as well as
for the classical model  \cite{ScL:JPA03}. In case of only nearest-neighbor
exchange $J_1$ the classical ground-state manifold contains coplanar as well as non-coplanar
states and the magnetization curve is a simple straight line.

\begin{figure}[ht!]
\hspace{0cm}\includegraphics*[clip,width=1.0\columnwidth]{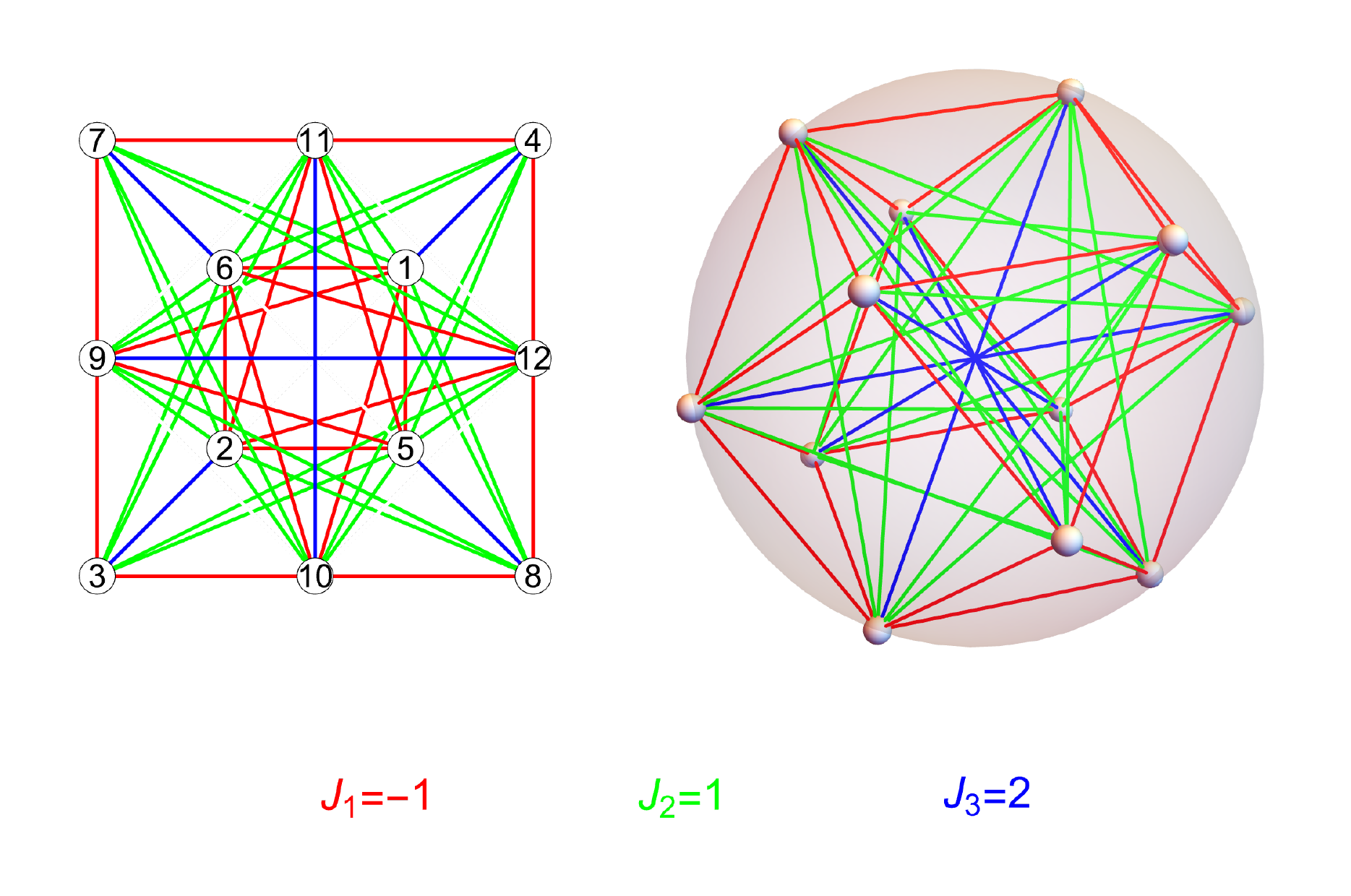}
\caption{Exchange matrix $\{J_{ij}\}$ leading to the cuboc state of the 12 spin model,
planar and three-dimensional representation.
}
\label{fig_model_i_TOY}
\end{figure}

For our purposes, we consider the inverse problem,
namely how to define suitable exchange bonds between $N=12$ spin sites so that the
resulting ground state forms a cuboctahedron in {\em spin space}.
Obviously, $N=12$ is the lowest number of spins with a cuboc ground state.
The details of the solution can be found in the Appendix.
As the result, we obtain the '12 spin model', i.~e., an exchange matrix $\{J_{ij}\}$
with three different exchange bonds $J_1=-1$, $J_2=1$, $J_3=2$.
It commutes with the natural 12-dimensional representation of ${\cal O}_h$,
i.e., it has  full ${\cal O}_h$ symmetry.
A graphical representation is given
in the left panel of Fig.~\ref{fig_model_i_TOY}.
It is also possible to
arrange the 12 spins on the vertices of the
cuboctahedron as it is shown in the right panel of
Fig.~\ref{fig_model_i_TOY}.
Note, however,
that the above choice of exchange bonds $J_1$, $J_2$, $J_3$ and the corresponding
arrangement of spins on the cuboctahedron is not unique.
An interchange of $J_1$ and $J_2$, i.e., a parameter set  $J_1=1$, $J_2=-1$, $J_3=2$
and a corresponding alternative arrangement of spins on the cuboctahedron
is equivalent. As a result, the
assignment of the cuboc ground state to either the cuboc1 or cuboc2 states
found for lattice models is not reasonable, since it depends on the arrangement
of the bonds.

\begin{figure}[ht!]
\hspace{0cm}\includegraphics*[clip,width=1.3\columnwidth]{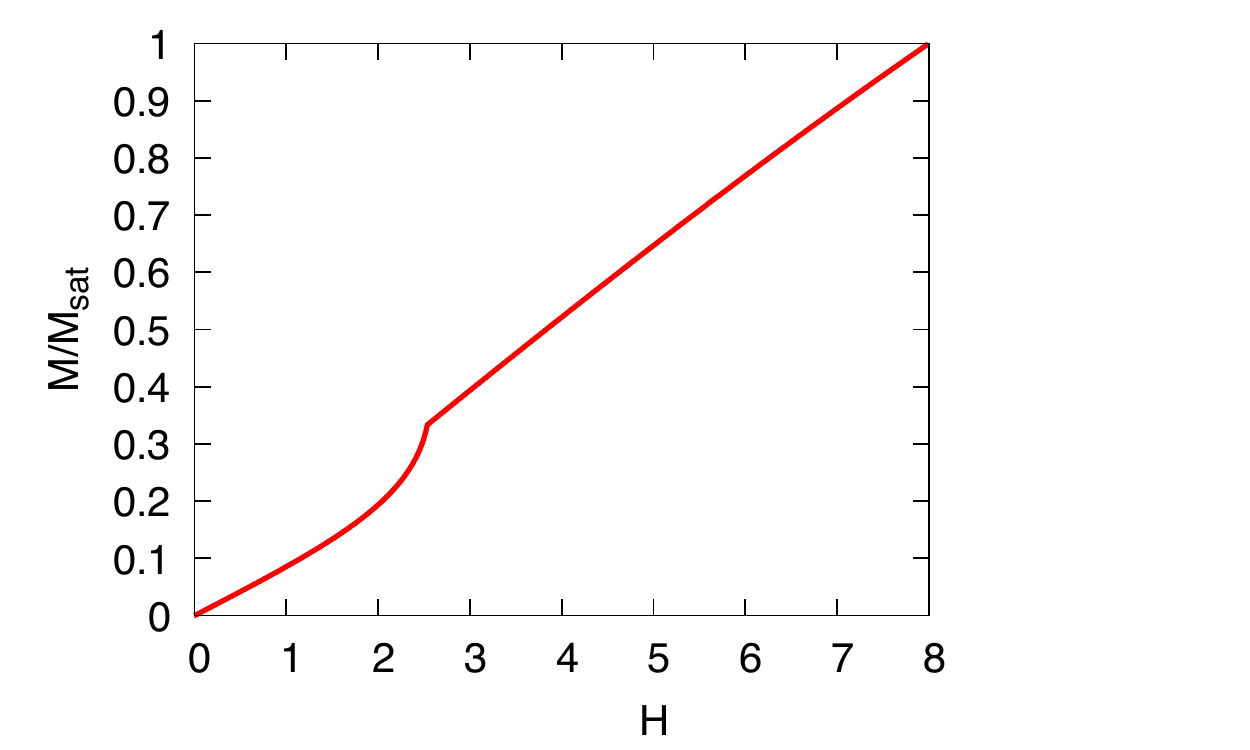}
\caption{Magnetization process of the $J_1$-$J_2$-$J_3$ 12 spin model
with a cuboc ground state. We observe a kink at $H=H_c=2.5359$
and $M=M_c=\frac{1}{3}M_{\rm sat}$.
 }
\label{fig_model_i_M_H}
\end{figure}

In what follows we present the main findings for the magnetization process of the
12 spin model.
The  magnetization process is shown  in Fig.~\ref{fig_model_i_M_H}.
The most striking feature of the magnetization curve is
a continuous phase transition at $H_c=2(3-\sqrt{3}) \approx 2.5359$ characterized by
a significant kink in $M(H)$ at $H_c$.
Moreover, the magnetization curve $M(H)$ is
nonlinear and the ground state is non-coplanar
in the whole region $0\le H < H_{\rm sat}=8$.
The magnetization at the transition is
$M(H_c) =M_{\rm sat}/3$.

\begin{figure}[ht!]
\hspace*{-0.1cm}\includegraphics*[clip,width=1.05\columnwidth]{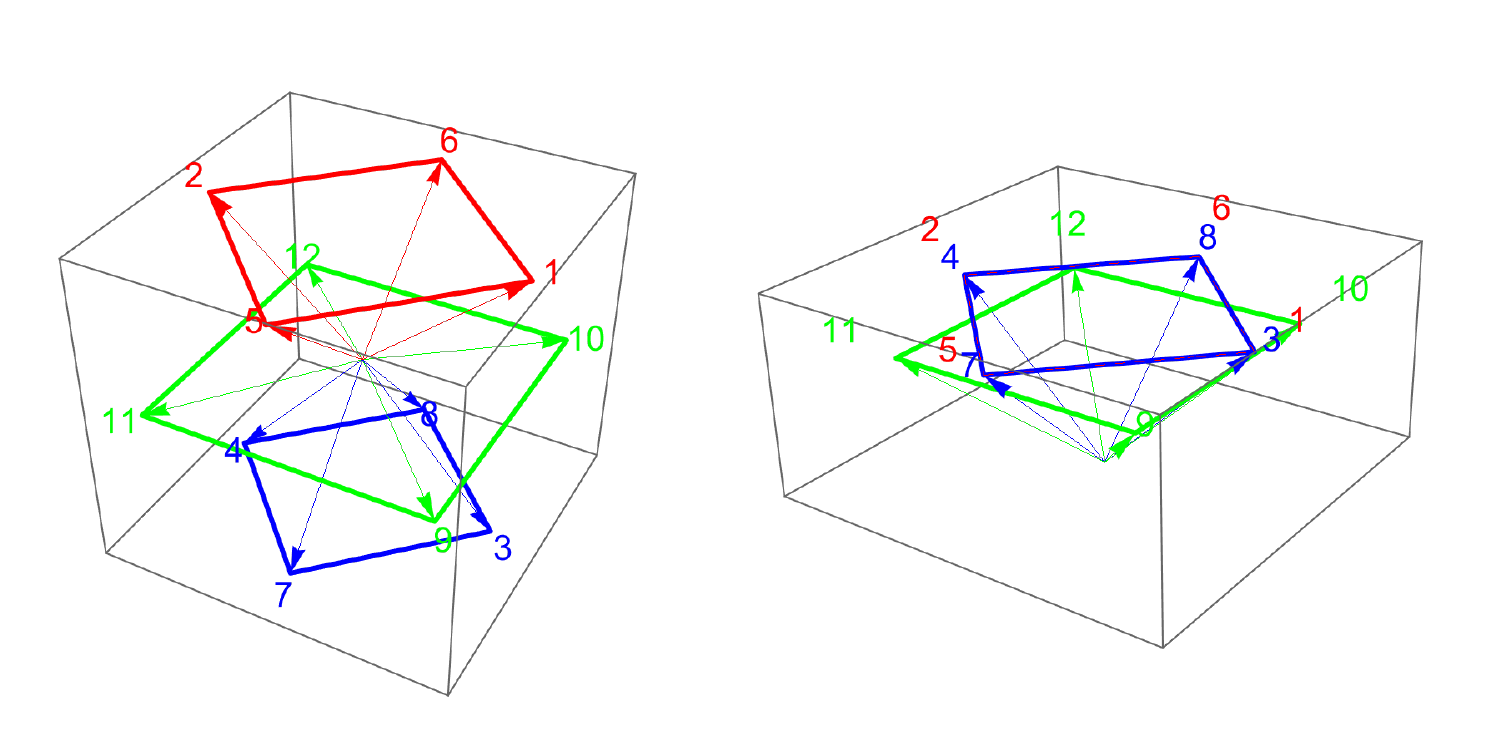}
\caption{Representation of the  ground states of the 12 spin model at
low and high magnetic fields; left $H=1.5<H_c$, right
$H=5>H_c$.
}
\label{fig_model_i_GS}
\end{figure}

\begin{figure}[ht!]
\hspace{-1cm}\includegraphics*[clip,width=0.9\columnwidth]{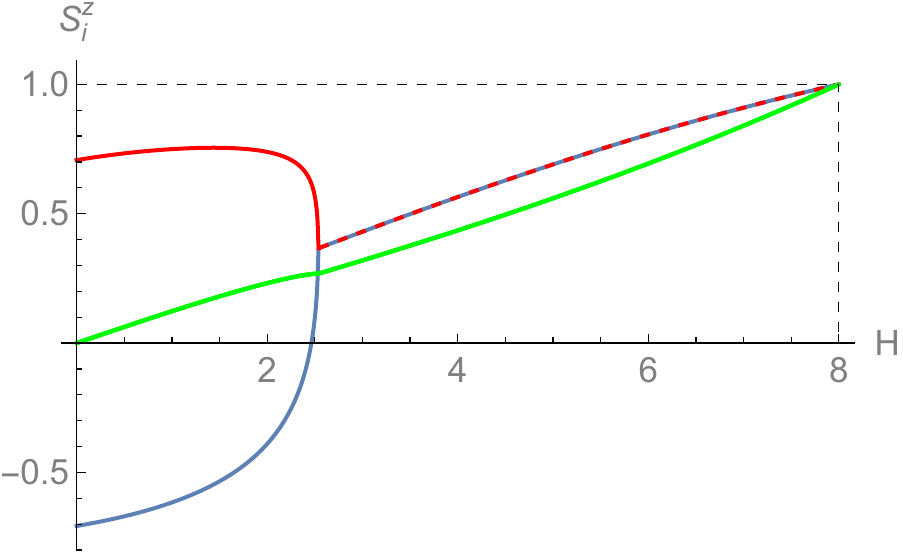}
\caption{The $z$-components of the spin vectors of the 12 spin  model
in dependence on the magnetic field.
The colors of the lines correspond to the colors used for the arrows in
Fig.~\ref{fig_model_i_GS}.
 }
\label{TOYSz}
\end{figure}

For  the low-field phase at  $H<H_c$ we use our semi-analytical approach.
The ground state is a  deformed cuboc state which consists of three groups of spins
forming squares
with respective polar angles $\theta_i,\;i=1,2,3$. Its energy is given by
\begin{eqnarray}\nonumber
&&  E(\theta_1,\theta_2,\theta_3;H)= \nonumber \\
&& \quad \frac{1}{3} \bigg\{\cos (\theta _1) \big[4 \cos (\theta _2)-H\big]-H \cos(\theta _2)
   -H \cos (\theta _3)  \nonumber\\
&& \qquad -\sin ^2(\theta _3)-2 \sin
   (\theta _1) \sin (\theta _2) \nonumber \\
   && \qquad -2 \sqrt{2}\big[ \sin (\theta _1) \sin
   (\theta _3)+ \sin (\theta _2) \sin (\theta _3) \big]
 \nonumber \\
&& \qquad -\cos
   ^2(\theta _1)-\cos ^2(\theta _2)  +\cos
^2(\theta_3)\bigg \}
   \; .
   \label{energyphaseI}
\end{eqnarray}
The polar angles are found  numerically by searching for the minimum of
$E(\theta_1,\theta_2,\theta_3;H)$
for fixed $H$.
We show the deformed cuboc ground state
in Fig.~\ref{fig_model_i_GS}, left panel, where the  three groups of spins
are presented as red, blue and green arrows pointing to three
squares, where the green square is twisted by  $45^\circ$.
The respective $z$-components as a function of the field $H$ are shown in
Fig.~\ref{TOYSz}. Interestingly, the  $z$-component $s^z_{\rm red}$ of the red group of spins
shows a non-monotonic behavior. After the expected increase of $s^z_{\rm red}$
at low fields it rapidly drops down as $H \to H_c$, even with an infinite slope
at $H = H_c$. We notice that also the  $z$-component of the blue group of spins
exhibits  an infinite slope as approaching the transition from below.
On the other hand, the slope of the magnetization at $H = H_c$ remains
finite:
$\lim_{H\uparrow H_c}\frac{\partial M}{\partial H}
 =\frac{1}{47} \left(21+10\sqrt{3}\right)\approx 0.81533$.

\begin{figure}[ht!]
\hspace{-1cm}\includegraphics*[clip,width=0.9\columnwidth]{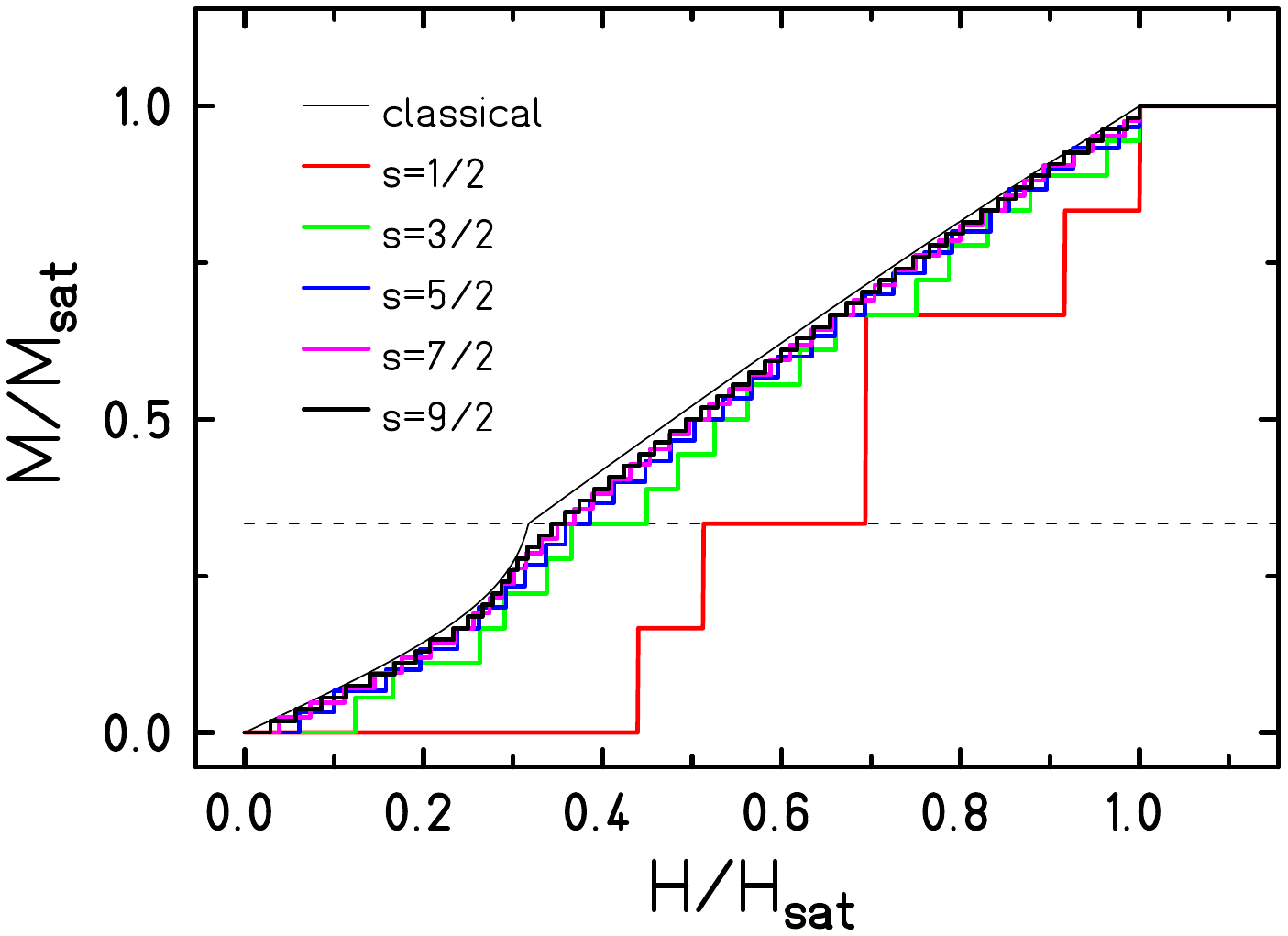}
\caption{Magnetization curves for the quantum 12 spin model and various
spin quantum numbers $s=\frac{1}{2},\ldots, \frac{9}{2}$.
 }
\label{fig_cuboc-s-M-H}
\end{figure}

\begin{figure}[ht!]
\hspace{-1cm}\includegraphics*[clip,width=0.9\columnwidth]{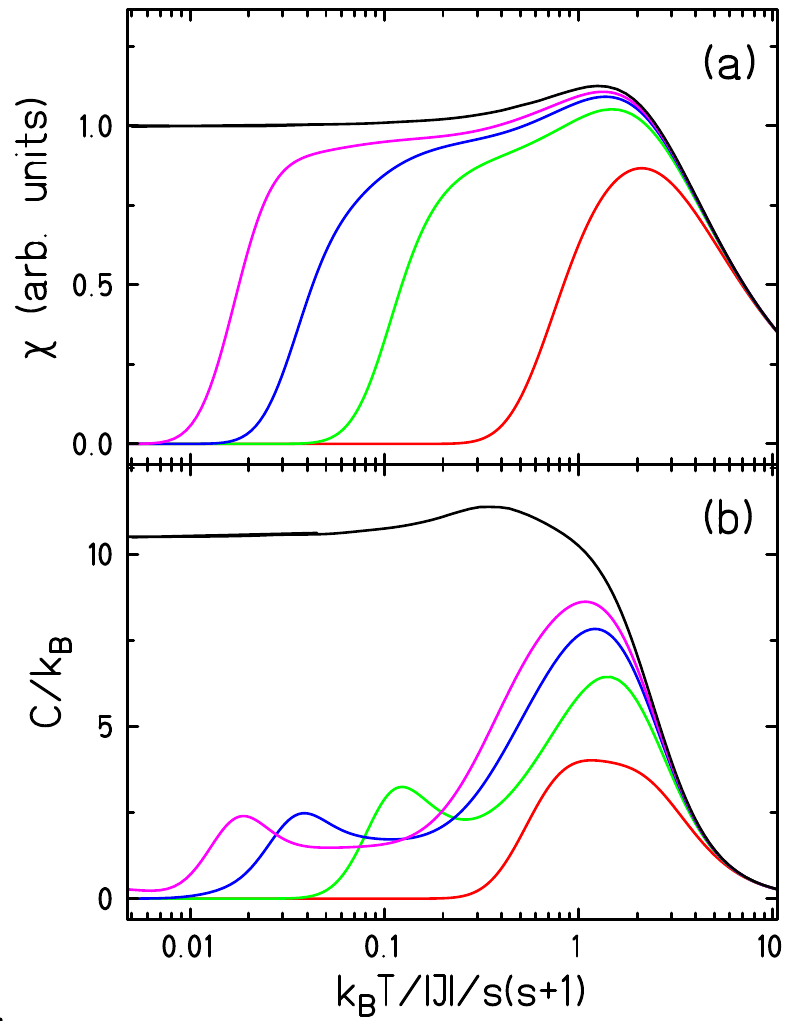}
\caption{Susceptibility $\chi$ and specific heat $C$ for the quantum 12 spin model
and various spin quantum numbers $s=\frac{1}{2},\ldots, \frac{7}{2}$ as function
of the scaled temperature $\frac{k_B\,T}{|J| s(s+1)}$.
The color codes are identical to those in Fig.~\ref{fig_cuboc-s-M-H}.
 }
\label{fig_cuboc-s-Chi-C-T}
\end{figure}

For the phase at  $H>H_c$ we can present a full analytical description.
The ground state consists now of two groups of spins
forming squares
with respective
polar angles $\theta_i,\;i=1,2$.
The energy of this state is given by
\begin{eqnarray}\nonumber
\label{energy_phaseII}
&& E(\theta_1,\theta_2;H) =
\frac{1}{6} H \big[ 2 \cos (\theta _1)+\cos (\theta _2)\big]
\nonumber \\
&& \quad +\frac{1}{6}
   \big[-8 \sqrt{2} \sin (\theta _1) \sin (\theta _2)+4 \cos (2 \theta_1) \nonumber \\
&& \qquad \quad +2 \cos (2 \theta _2)\big]
   \;.
\end{eqnarray}
The ground state satisfies
$
\frac{\partial E}{\partial\theta_1}= \frac{\partial E}{\partial\theta_2}=0
$
which can be solved for a parametric representation in terms of $z\equiv
\cos\theta_2$:
\begin{eqnarray}
\label{parasola}
  \cos \theta_1 &=& \frac{\sqrt{2} z}{\sqrt{z^2+1}}\;, \\
  \label{parasolb}
 H &=& 4 z \left(\frac{\sqrt{2}}{\sqrt{z^2+1}}+1\right)\;,\\
 \label{parasolc}
 M &=&\frac{1}{3} \left(\frac{2 \sqrt{2} z}{\sqrt{z^2+1}}+z\right)
 \;.
\end{eqnarray}
We mention,
that these equations, in principle,  allow to find an analytical expression for
$M(H)$ by eliminating the parameter $z$,
however, in form of the cumbersome expressions of the roots of a 4th order
polynomial.
The spin configuration is
a 'double umbrella' state, see Fig.~\ref{fig_model_i_GS},
right panel, where the  two groups of spins
are presented as blue and green arrows pointing to two
squares twisted by  $45^\circ$.
The blue group is the merging of the former blue and red groups of spins
of the  low-field phase, i.e., it contains 8 spins.
As the magnetic field approaches the saturation field,
both umbrellas converge toward the north pole.
If the magnetic field approaches $H_c$ from above we have
$\lim_{H\downarrow H_c}\frac{\partial M}{\partial H}= (15+2 \sqrt{3})/142\approx  0.130029$.
This yields a jump of the susceptibility of $\Delta \chi \approx  -0.685301$ at $H=H_c$.

Finally, we consider the quantum-mechanical 12 spin model.
With the exchange interactions given in \figref{fig_model_i_TOY}
the ground state, i.~e., $T=0$ magnetization curve of
the quantum spin model can numerically be obtained
for spin quantum numbers up to $s=9/2$ using a
Lanczos procedure \cite{Lan:JRNBS50}. Figure \ref{fig_cuboc-s-M-H} shows
the respective magnetization curves for half-integer
spin quantum numbers. Apart from the case $s=1/2$
the quantum curves approach the classical curve closely
from below, and the larger $s$ is, the more so.
For the largest available spin of $s=9/2$,
even the non-linear increase into the kink is clearly visible.

For completeness, we also show the temperature dependence
of the magnetic susceptibility as well as of the heat capacity
for 12 spin model systems with single-spin quantum number of up
to $s=7/2$,
see Fig.~\ref{fig_cuboc-s-Chi-C-T}.
For small spin quantum numbers exact diagonalization was used,
otherwise the thermal quantities were calculated using the finite-temperature
Lanczos method \cite{JaP:PRB94}.
For the interested reader we like to point out that the susceptibility
approaches the classical limit much quicker with increasing $s$
than the heat capacity.

In the following sections we consider lattice models.
We will see that features observed for the 12 spin model are also present for
extended systems. Thus this model may serve as a paradigm for a
classical magnetization curve starting from a cuboc ground ground, which
allows a comprehensive (semi-)analytical description of the magnetization
process.

\section{The Heisenberg model on the kagom\'{e} lattice with cuboc phases}
\label{sec-4}

\begin{figure}[ht!]
\hspace{-0.1cm}\includegraphics*[clip,width=1.0\columnwidth]{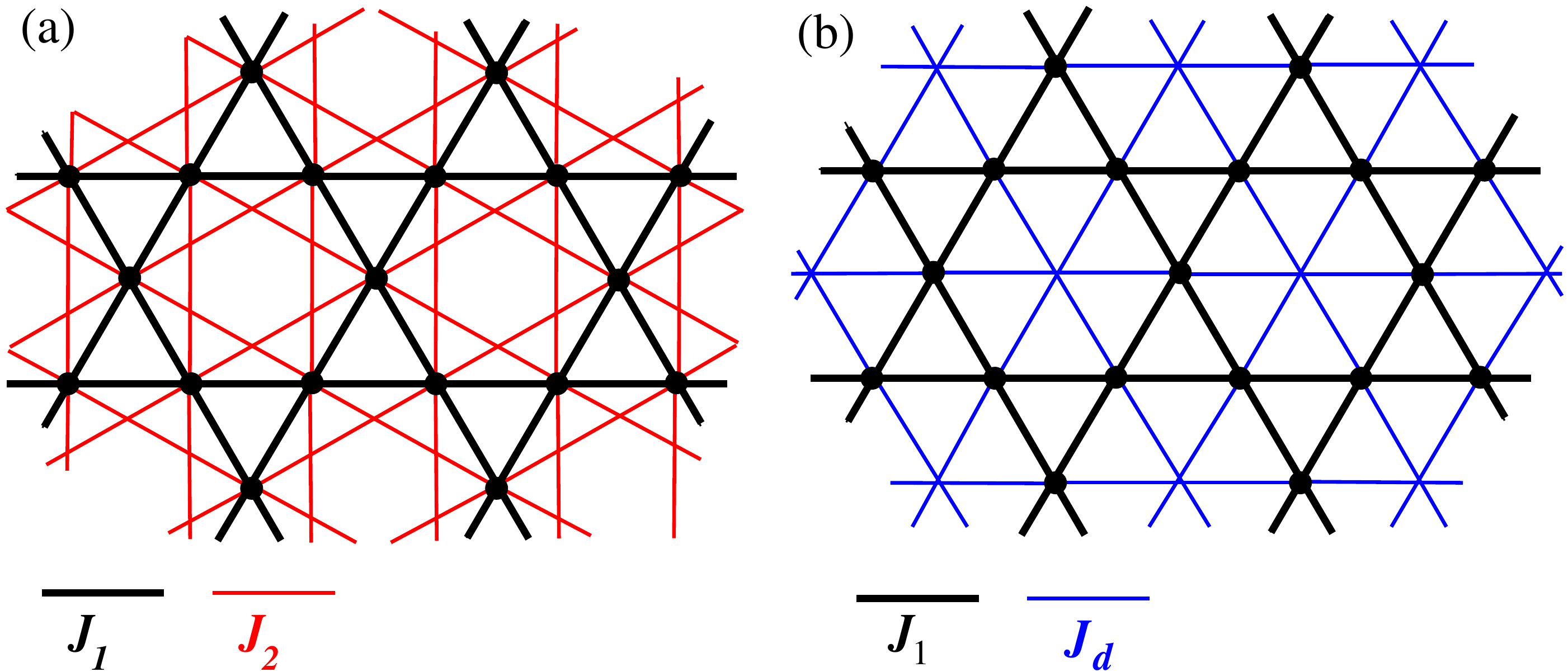}
\caption{Sketch  of the $J_1$-$J_2$ (a) and the  $J_1$-$J_d$ (b) model on the kagom\'{e} lattice.}
\label{fig_kago_J1_J2}
\end{figure}

\subsection{The $J_1$-$J_2$ model}
\label{sec-4.1}

As reported in Refs.~\cite{Domenge2005_cuboc2} and
\cite{Messio_2011_clas_GS}
a zero-field cuboc ground state of the type cuboc2 exists for
the $J_1$-$J_2$ Heisenberg model with ferromagnetic nearest-neighbor
bonds  $J_1$ and
antiferromagnetic  2nd neighbor bonds
$J_2$ (see Fig.~\ref{fig_kago_J1_J2} (a)), if $J_2/|J_1| \ge 1/3$.
We set the energy scale by choosing $J_1=-1$.
The saturation field is then given by $H_{\rm sat}=6(J_2-1/3)$.
In what follows, we will demonstrate that the  general features
of the magnetization process of the $J_1$-$J_2$ kagom\'{e} model are quite similar to
those of the 12 spin model. In particular, there are two phases with three groups of spins below a
critical field $H_c$ and with two groups of spins for $H>H_c$,
the magnetization curve $M(H)$ is
nonlinear and the ground state is noncoplanar
in the whole region $0\le H < H_{\rm sat}$.

\begin{figure}[ht!]
\includegraphics*[clip,width=1.1\columnwidth]{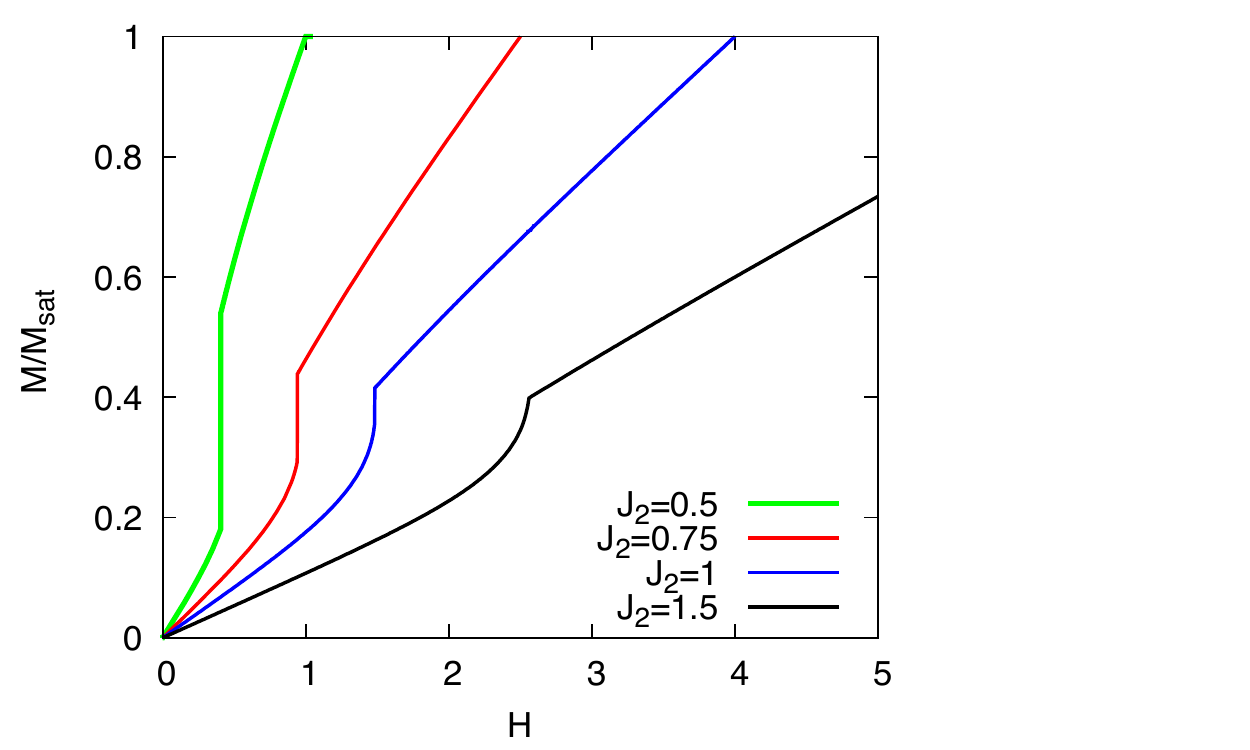}
\caption{Magnetization curves of the $J_1$-$J_2$ Heisenberg model on
the kagom\'{e} lattice for $J_1=-1$ and various values of $J_2$ (see legend).
 }
\label{fig_m_H_kago_J1_J2}
\end{figure}

\begin{figure}[ht!]
\hspace{-0.1cm}\includegraphics*[clip,width=1.0\columnwidth]{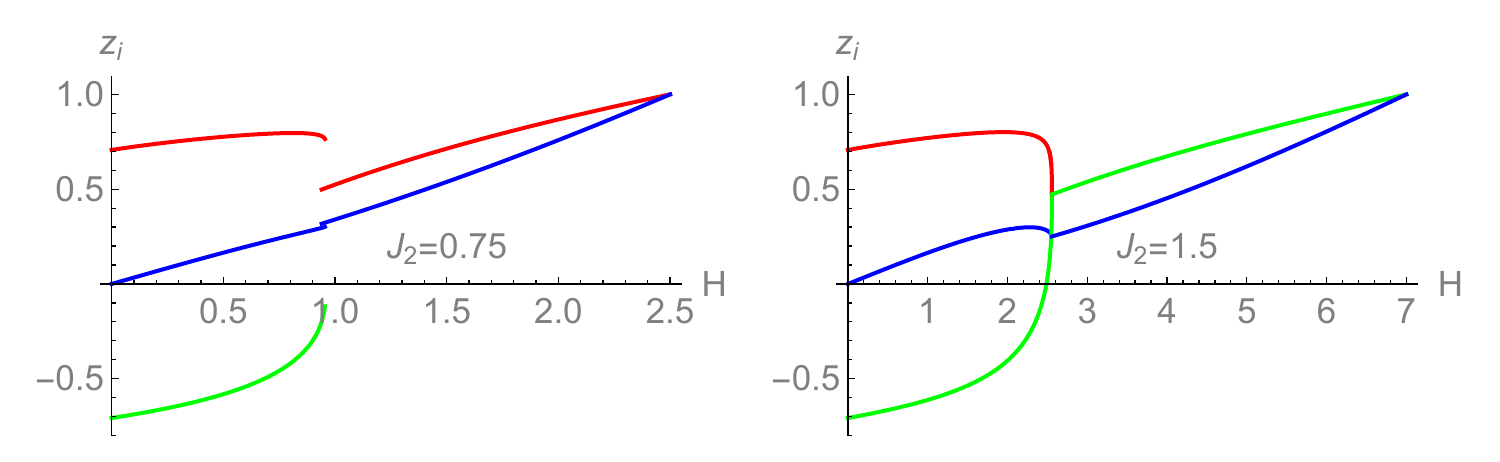}
\caption{The $z$-components of the spin vectors of
of the $J_1$-$J_2$ Heisenberg model on
the kagom\'{e} lattice for $J_1=-1$ and $J_2=0.75$ (left panel) and  $1.5$ (right panel)
in dependence on the magnetic field.}
\label{Sz_J1_J2}
\end{figure}

\begin{figure}[ht!]
\hspace{-0.4cm}\includegraphics*[clip,width=1.05\columnwidth]{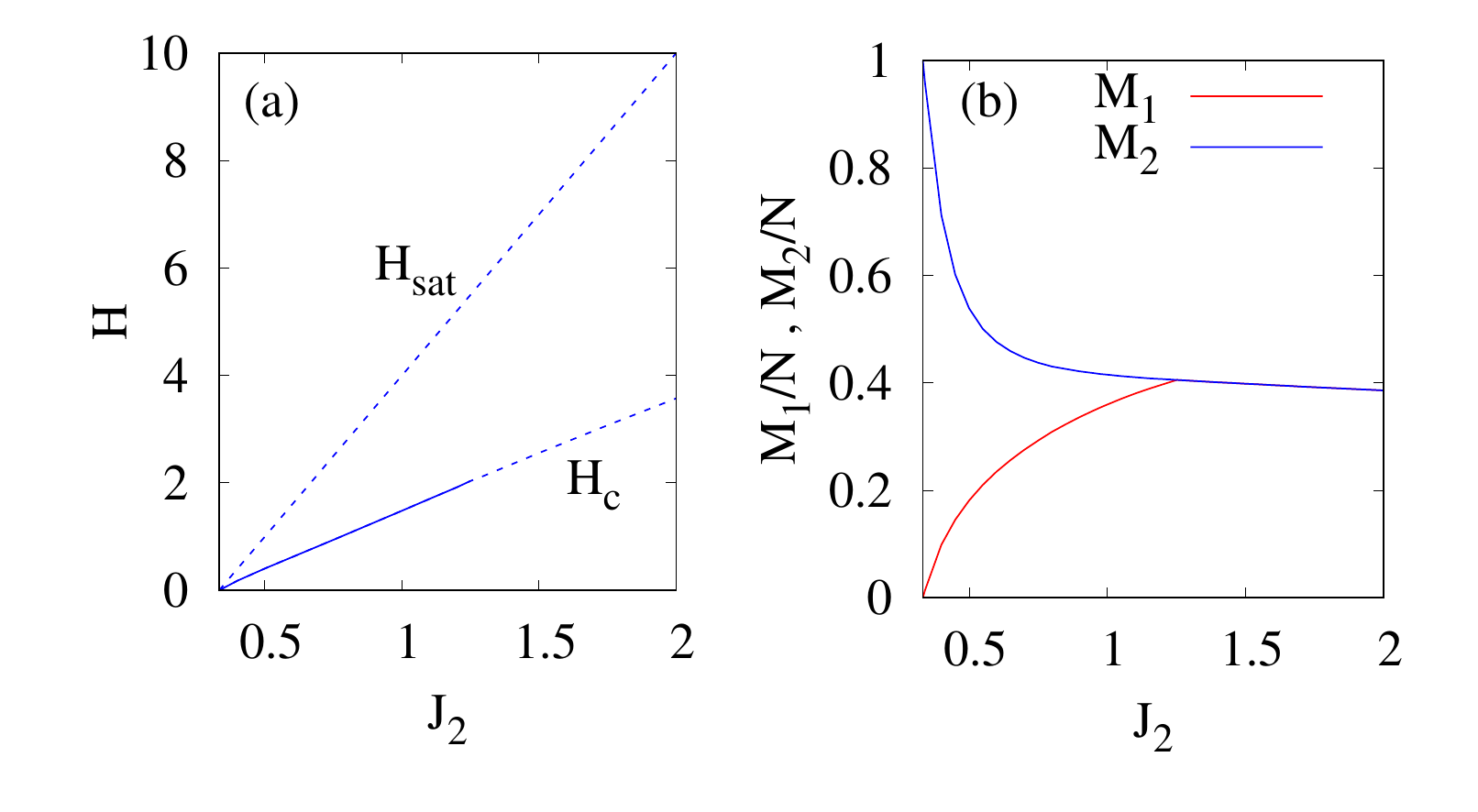}
\caption{(a) Phase diagram for  the $J_1$-$J_2$ kagom\'{e}  model (solid lines -
discontinuous transitions, dashed lines - continuous transitions).  (b)  Magnetizations $M_1$, $M_2$ at
the critical field $H_c$
in dependence on $J_2$ for  the $J_1$-$J_2$ kagom\'{e} model.
 }
\label{fig_Bc_M1M2_caseii}
\end{figure}

Similarly as in Sec.~\ref{sec-3} we can use the semi-analytical approach below
$H_c$,  and we can provide
a full analytical description above $H_c$  in form of a $J_2$-dependent parametric
representation  similar to  Eqs.~(\ref{parasola}) - (\ref{parasolc}).
We show a few magnetization curves in Fig.~\ref{fig_m_H_kago_J1_J2}.
The spin configurations of the two phases below and above $H_c$
are very similar to those shown in
Fig.~\ref{fig_model_i_GS}, i.e., there is a deformed cuboc state which consists of three groups of spins
at low field and a double umbrella state at high fields.

In contrast to the 12 spin model the phase transition at the critical
field $H_c$
is either continuous or discontinuous dependending on the magnitude of
$J_2$. For $J_2 < J_m \approx 1.25221$ there is a jump at $H_c$ in the
magnetization curve, whereas for
$J_2 \ge J_m$ the $M(H)$ curve exhibits a kink  at $H_c$.
We mention that for  $J_2=J_m$ the differential susceptibility diverges as
approaching $H_c$ according to
$dM/dH \sim (H_c-H)^{-1/2}$.
It should be noted, however, that this relationship
as well as the previous remarks do not represent rigorous results,
but rather were obtained by analyzing numerical data from the semi-analytical approach.
In Fig.~\ref{fig_m_H_kago_J1_J2} we
notice the jump for $J_2=0.5$, $0.75$, and $1.0$, but only a kink
for $J_2=1.5$.
In Fig.~\ref{Sz_J1_J2}
we show the $z$-components of the spin vectors  as a function of the field $H$
for two representative parameter sets $J_2=0.75$ and $1.5$.
For $J_2=1.5$ the striking analogy to Fig.~\ref{TOYSz} is  obvious.
But also for $J_2=0.75$ the same two phases are present, however, with an
abrupt change of the spin configuration at the transition point $H_c$.
Finally
we present in
Fig.~\ref{fig_Bc_M1M2_caseii} the saturation field $H_{\rm sat}$, the critical
field $H_c$ as well as the
magnetizations  $M_1$ and $M_2$ at $H_c$ as functions of
$J_2$.

\subsection{The $J_1$-$J_d$ model}
\label{sec-4.2}

\begin{figure*}[ht!]
\includegraphics*[clip,width=2.0\columnwidth]{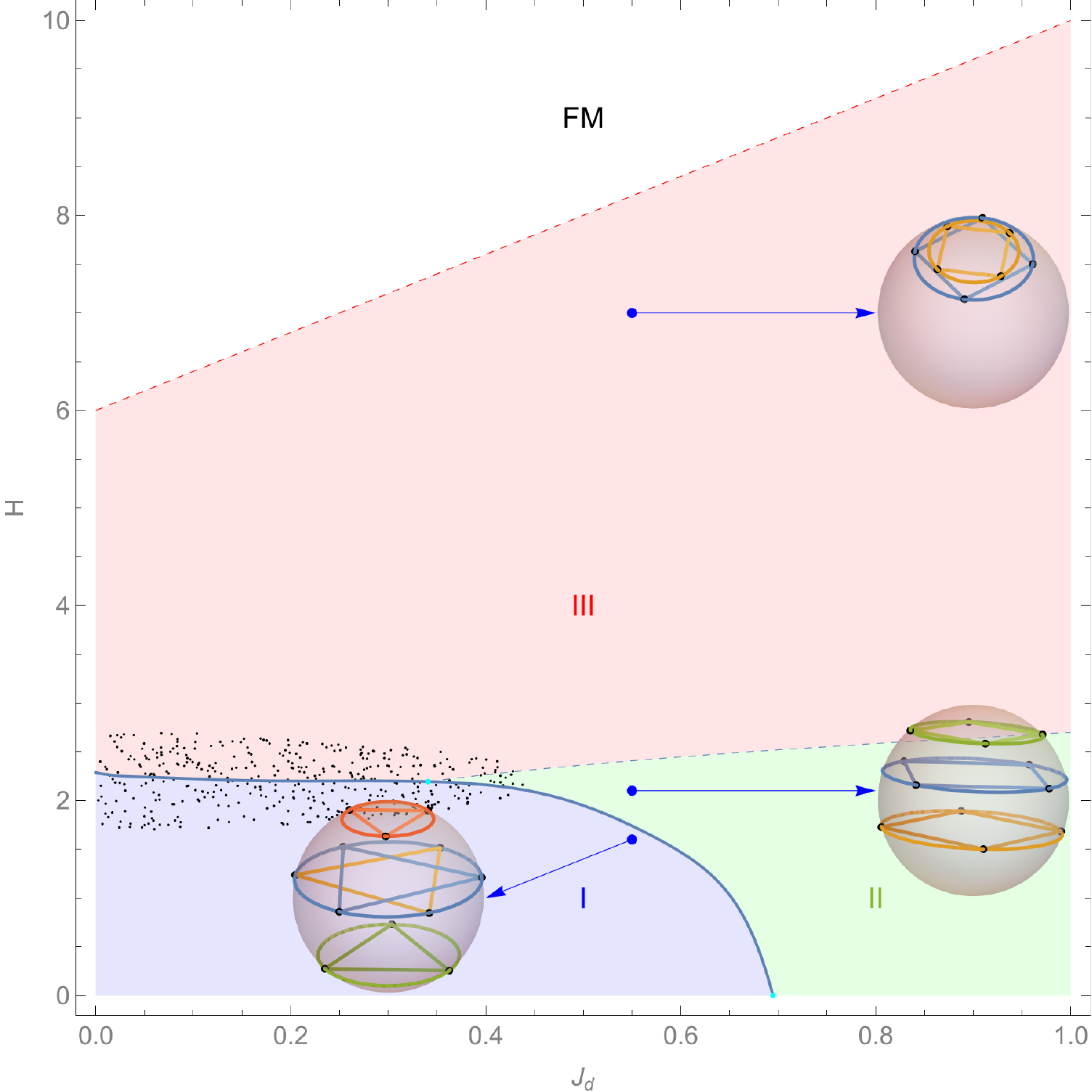}
\caption{
Phase diagram of the $J_1$-$J_d$ kagom\'{e} model setting $J_1=1$. The horizontal axis
represents $0\le J_d\le 1$ and the vertical axis the magnetic field $0\le H \le 10$.
We observe three phases denoted by ${\rm I}$, ${\rm II}$ and ${\rm III}$ with typical common origin plots
inserted.
Solid curves indicate discontinuous phase transitions and dashed curves continuous ones.
In the dotted region, the phases and their boundaries are
uncertain due to finite-size effects.
}
\label{fig_phadia_model_iii}
\end{figure*}

According to Refs.~\cite{janson2008modified,janson2009intrinsic} and
\cite{Messio_2011_clas_GS}
another zero-field cuboc ground state, now of the type cuboc1, exists for
the kagom\'{e} $J_1$-$J_d$ Heisenberg model with antiferromagnetic nearest-neighbor
bonds  $J_1$ and
antiferromagnetic  3rd neighbor bonds
$J_d$ along the diagonals of the
hexagons, see  Fig.~\ref{fig_kago_J1_J2}(b).
Some results for the magnetization curves of these systems
have already been published in Ref.~\cite{Zhito_M_H_cuboc},
in particular for $J_d=0.1\,J_1$ and $J_d=0.5\,J_1$.
We were able to confirm these results, but have to make some restrictions
due to finite-size effects for the case $J_d=0.1$, see below.

In what follows we set the energy scale by choosing $J_1=1$.
The saturation field is then given by
$H_{\rm sat}= 6+4\,J_d$.
Again we will observe that the magnetization process exhibits
similarities to the 12 spin model discussed in Sec.~\ref{sec-3}.
The influence of the magnetic field on the ground states of the kagone lattice
is summarized in the phase diagram shown in Fig.~\ref{fig_phadia_model_iii}.

Additional to the ferromagnetic phase FM and the cuboc1 phase for $H=0$ we encounter three
non-coplanar phases denoted by I, II and III. Phase I and II prevail at low fields and
represent two different types of deformed cuboc1 states,
each consisting of three groups of spins with constant $z$-components.
As for the previous models, in the high-field phase III just before saturation
a full analytical description is possible in form of a $J_d$-dependent parametric
representation  similar to  Eqs.~(\ref{parasola}) -(\ref{parasolc}).
We omit the explicit form because it is too complicated.

\begin{figure}[ht!]
\includegraphics*[clip,width=1.0\columnwidth]{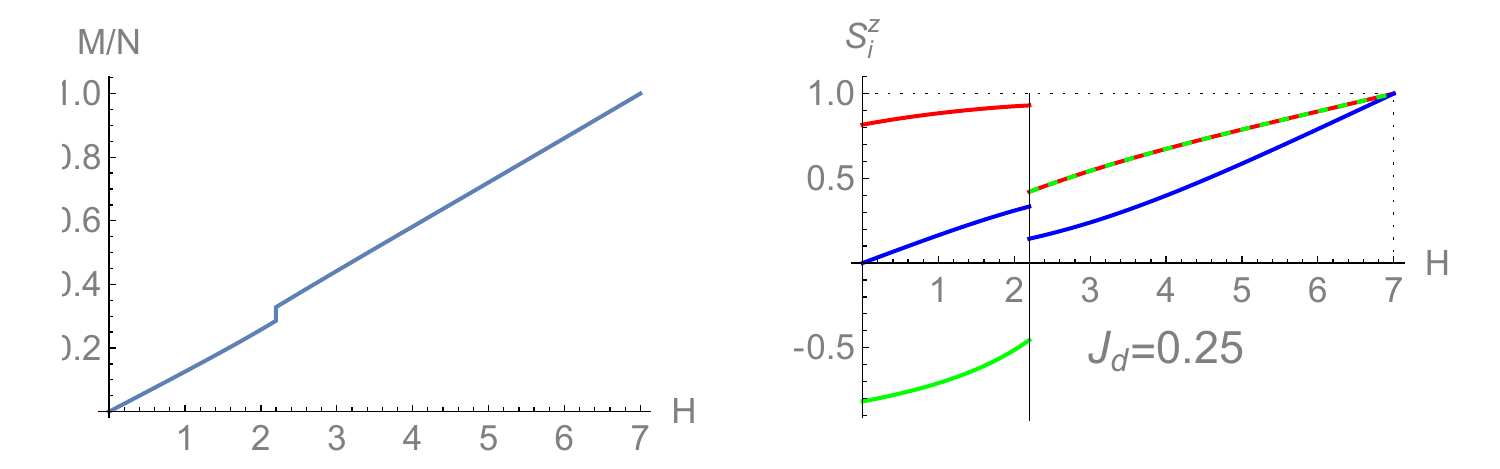}
\caption{Magnetization curve  of the $J_1$-$J_d$ Heisenberg model on
the kagom\'{e} lattice  for $J_1=1$ and $J_d=0.25$ (left) and  corresponding $z$-components of the
spin vectors (right).}
\label{fig_m_Sz_H_kago_J1_Jd_025}
\end{figure}

\begin{figure}[ht!]
\includegraphics*[clip,width=1.0\columnwidth]{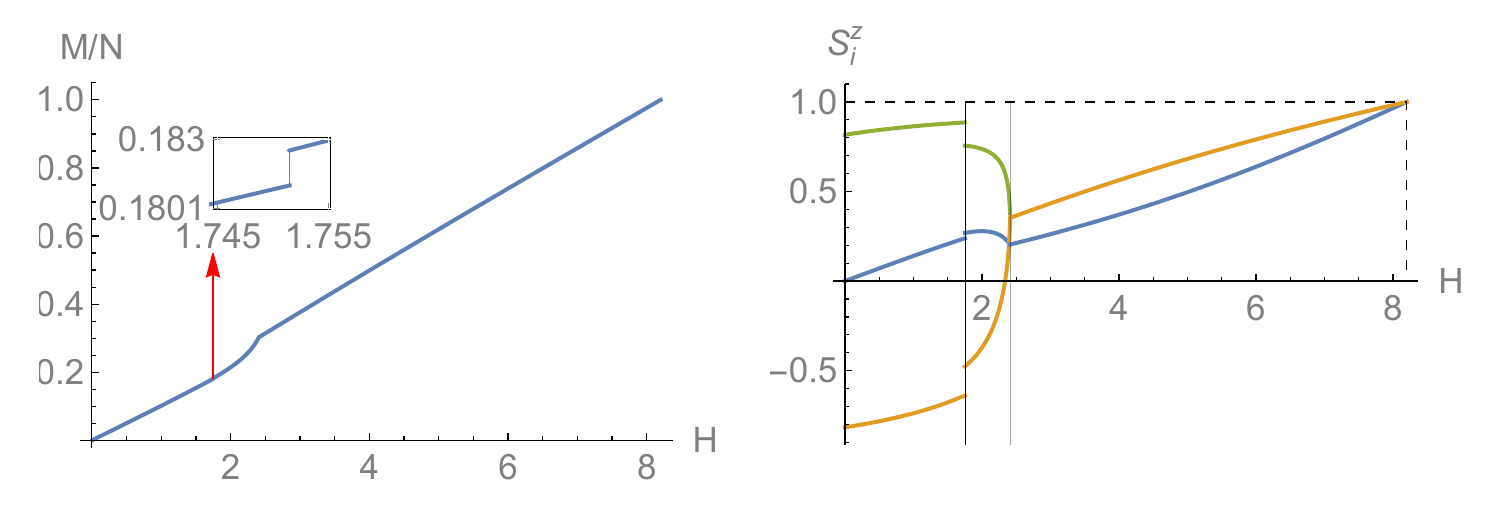}
\caption{Magnetization curve (left) and corresponding $z$-components of the
spin vectors (right) of the $J_1$-$J_d$ Heisenberg model on
the kagom\'{e} lattice for $J_1=1$ and  $J_d=0.55$.
The inset enlarges the scale in order to show the discontinuous phase transition
at $H_c=1.7515$.
}
\label{fig_m_Sz_H_kago_J1_Jd_055}
\end{figure}

\begin{figure}[ht!]
\includegraphics*[clip,width=1.0\columnwidth]{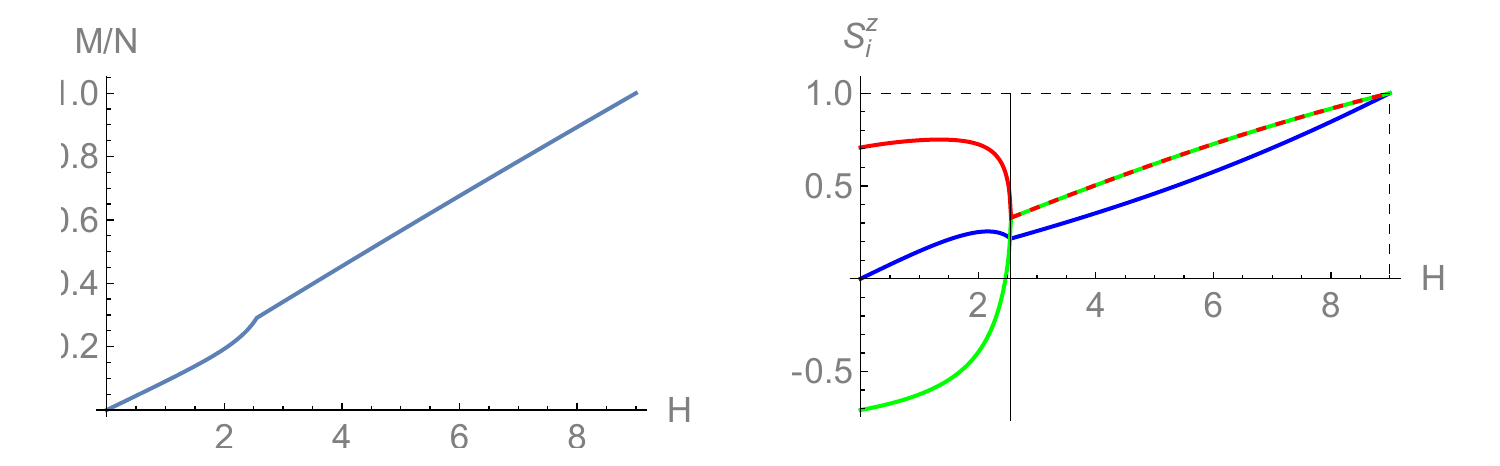}
\caption{Magnetization curve (left) and corresponding $z$-components of the
spin vectors (right) of the $J_1$-$J_d$ Heisenberg model on
the kagom\'{e} lattice for $J_1=1$ and  $J_d=0.75$.
}
\label{fig_m_Sz_H_kago_J1_Jd_075}
\end{figure}

According to the phase diagram there are three kinds of magnetization processes
with different transitions between the various phases.
\begin{enumerate}
  \item  For smaller values $J_d \lesssim 0.341$  there is one discontinuous
transition  I -- III, see the example  for $J_d=0.25$
in Fig.~\ref{fig_m_Sz_H_kago_J1_Jd_025} showing the magnetization curve and the
$z$-components of the spins in the ground state depending on $H$.
  \item  For intermediate values $0.341 \lesssim J_d \lesssim 0.694$
there is one discontinuous transition I -- II and one continuous one II - III,
which are located close to each other.  Corresponding data for
$J_d=0.55$ are shown in Fig.~\ref{fig_m_Sz_H_kago_J1_Jd_055}.
  \item At larger values $J_d \gtrsim 0.694$
there is only one continuous transition II -- III, see
Fig.~\ref{fig_m_Sz_H_kago_J1_Jd_075}, and the overall behavior of the
magnetization and  the $z$-components corresponds to that of the 12 spin model.
\end{enumerate}

\begin{figure}[ht!]
\includegraphics*[clip,width=1.1\columnwidth]{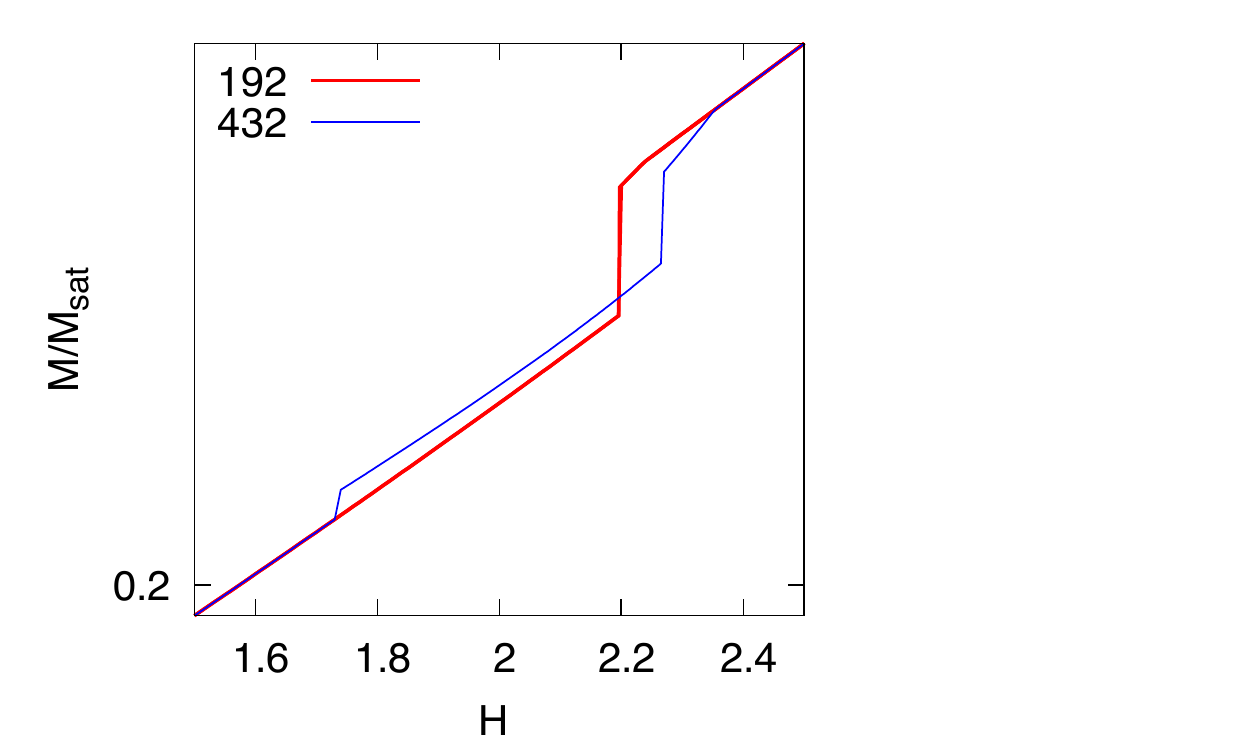}
\caption{Magnetization curve of the $J_1$-$J_d$ Heisenberg model on
the kagom\'{e} lattice for $J_1=1$ and  $J_d=0.25$ and the interval $1.5<H<2.5$.
The number of spins $N=3L^3$ is chosen as $N=192$ and $N=432$.
This illustrates the occurrence of finite-size
effects in a certain region for  $J_d\lesssim 0.45$.
}
\label{fig_m_Sz_H_kago_J1_Jd_025_fze}
\end{figure}

For a certain region with  $J_d \lesssim 0.45$ our numerical data are not entirely conclusive
since the phase transitions depend on the number $N$ of spins considered in the model,
see Figure \ref{fig_m_Sz_H_kago_J1_Jd_025_fze}.
These finite-size effects are possibly due to tiny energy differences
between the phases that occur at small values of $J_d$.

\section{The Heisenberg model on the square-kagom\'{e} lattice with cuboc phases}
\label{sec-5}

\begin{figure}[ht!]
\centering
\includegraphics*[clip,width=0.8\columnwidth]{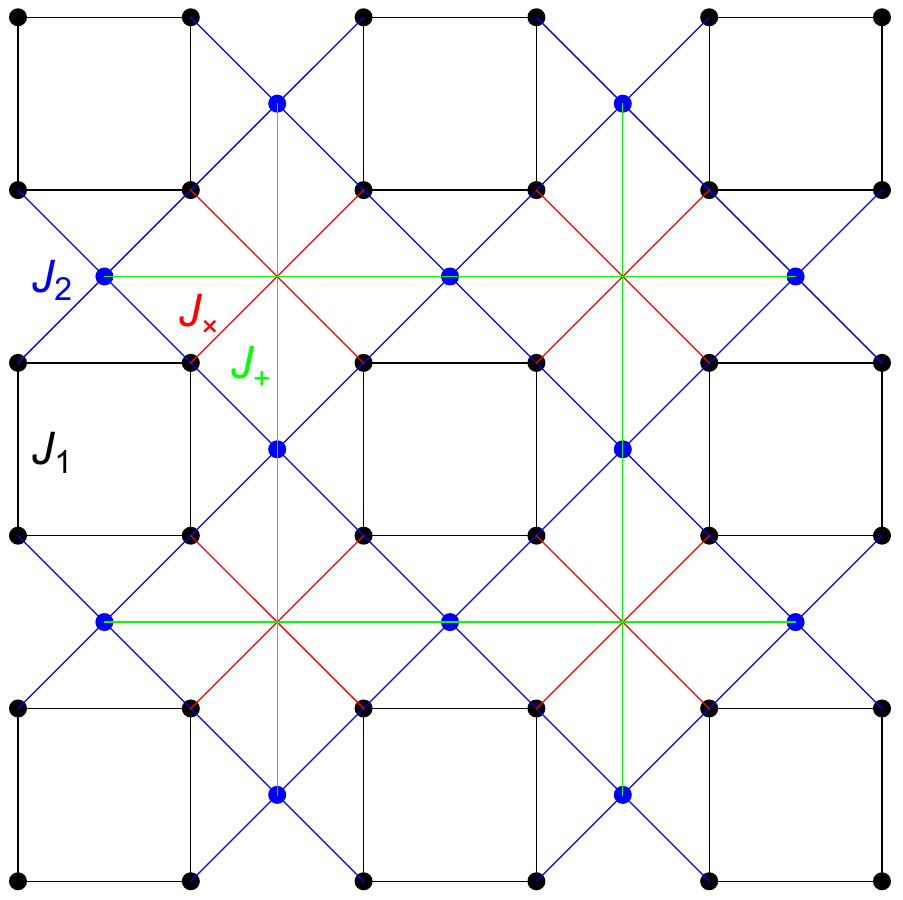}
\caption{ The square-kagom\'{e} $J_1$-$J_2$ model with cross-plaquette interactions
$J_+$ and $J_{\rm x}$. For the sake of simplicity, we set $J_+=J_{\rm x}=J_3$.
The sites A (black dots) form the squares and the sites B (blue dots)
sit in the middle of the bow-ties connecting the squares.
}
\label{fig_MOD}
\end{figure}

The magnetization process as well as the relevant theoretical approaches for
the Heisenberg model on the square-kagom\'{e} lattice with cuboc phase have been
discussed in detail very recently in Ref.~\cite{M_H_cuboc_I}.
However, in order to obtain a comprehensive picture of the magnetization processes
of classical Heisenberg systems with zero-field cuboctahedral ground state,
it seems useful to recapitulate the main features of the magnetization process
for square-kagom\'{e} models and to compare them with the models discussed in the previous sections.

A striking contrast to the kagom\'{e} model is
the existence of two non-equivalent sites
A (forming the squares) and B (sitting at the center of the bow ties connecting the squares)
as well as two non-equivalent  nearest-neighbor bonds $J_1$ and
$J_2$.
A comprehensive study of the zero-field ground-state phase diagram of the classical
square-kagom\'{e} spin model including cross-plaquette interactions has been
presented recently in Ref.~\cite{squago_clas_2023}.
The corresponding model is depicted in Fig.~\ref{fig_MOD}.

There are two non-equivalent  cross-plaquette interactions $J_+$ and $J_{\rm x}$.
However,   the zero-field cuboc ground-state phase is present in the entire
parameter region $J_+,J_{\rm x}>0$ independent of the magnitudes of $J_+$ and $J_{\rm
x}$, see Fig. 5 in Ref.~\cite{squago_clas_2023}.
For the sake of simplicity, therefore we consider the symmetric case
$J_+=J_{\rm x}=:J_3$, only.

\begin{figure}[ht!]
\centering
\includegraphics*[clip,width=1.05\columnwidth]{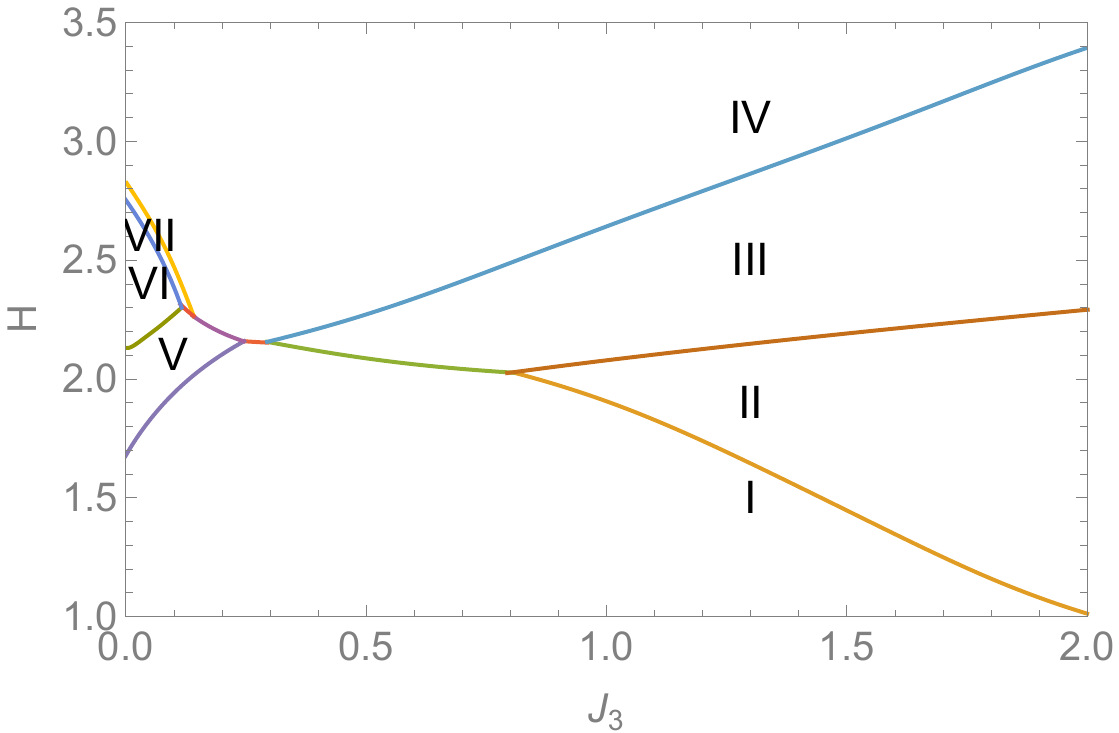}
\caption{Phase diagram for  $J_1=J_2=1$. The ferromagnetic
phase for $H\ge H_{\rm sat}$ is not shown.
}
\label{fig_PD1}
\end{figure}

\begin{figure}[ht!]
\centering
\includegraphics*[clip,width=1.0\columnwidth]{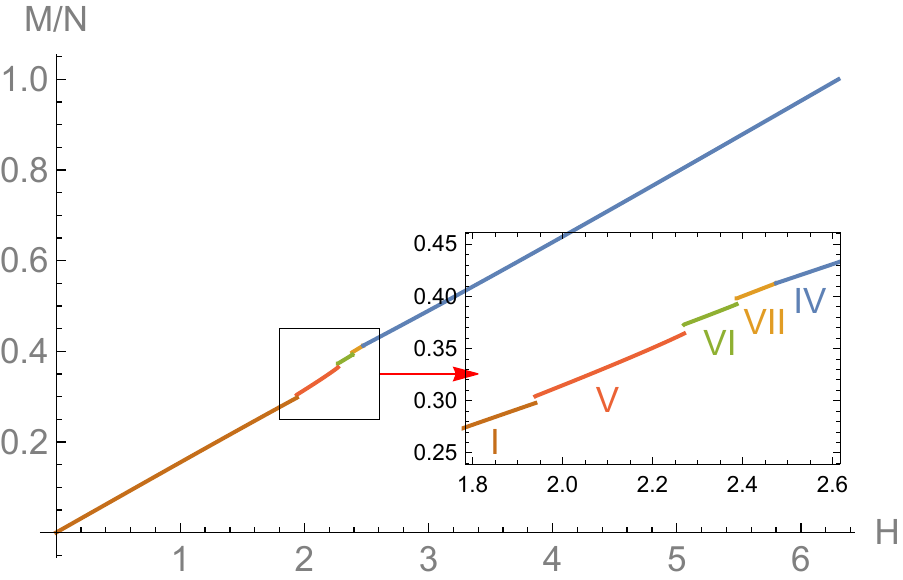}
\caption{Magnetization curve of the square-kagom\'{e} for $J_1=J_2=1$, $J_3=0.1$.
The inset enlarges the scale to make the phase transitions more clearly visible.
 }
\label{fig_BM01}
\end{figure}

\subsection{The model with antiferromagnetic (AF) nearest-neighbor
bonds
}
\label{sec-5.1}
First we consider the square-kagom\'{e} model with $J_1=J_2=1$ and $J_3>0$.
From Ref.~\cite{squago_clas_2023} we know that the zero-field ground
state is a so-called cuboc1 state, where
all neighboring pairs of spins form angles of 120°.
The saturation field is given by
\begin{equation}\label{hsat_squago_AFM}
H_{\rm sat}=
 4+3J_3+\sqrt{J_3^2+4}
 \;.
\end{equation}

This model shows a rich variety of seven non-coplanar
phases as a function of $J_3$ and $H$, denoted by ${\rm I}$ to ${\rm VII}$, see Figure \ref{fig_PD1}.
Of the six possible types of magnetization curves,
we will only describe the case with small $J_3$ (e.g. for $J_3=0.1$).
Here we see three jumps and two kinks
corresponding to the phase transitions
${\rm I}\stackrel{{\mathcal J}}{\to}{\rm V}\stackrel{{\mathcal J}}{\to}{\rm VI}\stackrel{{\mathcal J}}
{\to}{\rm VII}\stackrel{{\mathcal K}}{\to}{\rm IV}\stackrel{{\mathcal K}}{\to}FM$,
see Fig.~\ref{fig_BM01}.
For more magnetization curves see \cite{squago_clas_2023}.

The phase ${\rm I}$, which passes into the cuboc1 phase for small $H$,
shows 14 spins (sublattices) in the common origin plot,
which are distributed over the north- and south pole of the unit sphere
and over three squares, each with constant $z$ components.
For $H\to 0$ the four spins of the center square merge in pairs
so that the remaining $12$ spins form the vertices of a cuboctahedron.
Phase ${\rm IV}$ forms the transition to the $FM$ ground state
and can be described analytically as a common-origin plot of two squares,
each with constant $z$ components, which are rotated by the angle 45°
with respect to each other.
Also noteworthy is the disordered phase ${\rm VII}$, which exists in a narrow strip
for small $J_3$ values, see Figure \ref{fig_PD1},
and can be considered a candidate for a spin liquid due to its high degeneracy.

\subsection{The model with ferro- and antiferromagnetic (FM/AF) nearest-neighbor
bonds
}
\label{sec-5.2}

Now we consider the square-kagom\'{e} model with $J_1=1$, $J_2=-1$ and $0 \le J_3  \le 2$.
From Ref.~\cite{squago_clas_2023} it is known that the zero-field ground
state is a so-called cuboc3 state, where
each neighboring pair of spins forms an angle of 120°
on the squares and an angle of 60° on the triangles.

The saturation field is given by
\begin{equation}\label{hsat_squago_FM}
H_{\rm sat}=
2 (1+J_{3}) \; , \; J_{3}\le 2
\;.
\end{equation}

\begin{figure}[ht!]
\centering
\includegraphics*[clip,width=1.0\columnwidth]{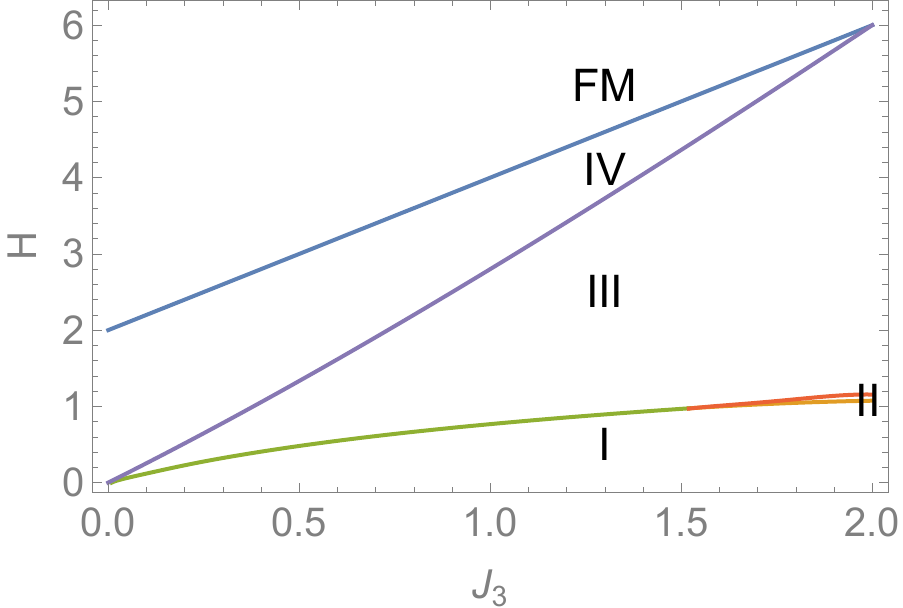}
\caption{Phase diagram for the FM/AF square kagom\'{e} with $J_1=-J_2=1$.
}
\label{fig_PD2}
\end{figure}

\begin{figure}[ht!]
\centering
\includegraphics*[clip,width=1.0\columnwidth]{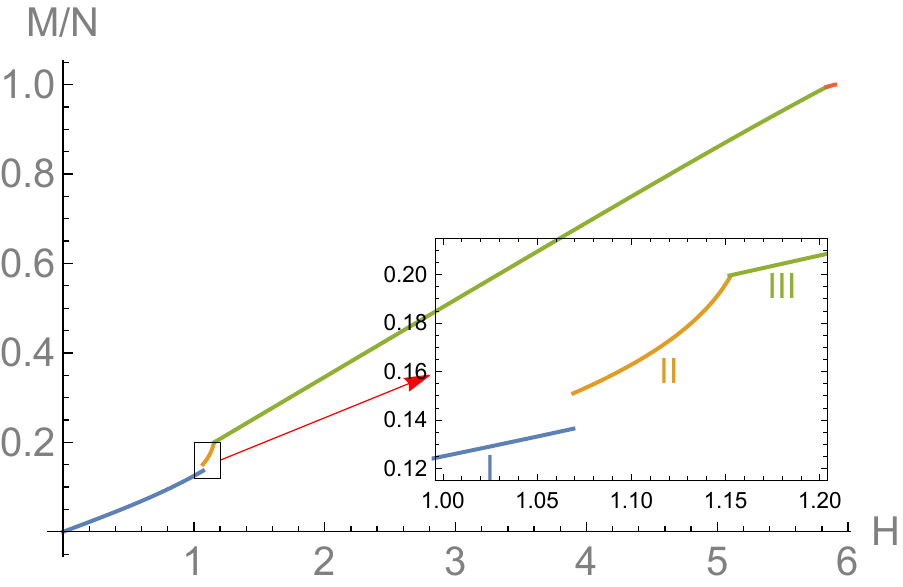}
\caption{Magnetization curve of the square-kagom\'{e} for $J_1=-J_2=1$, $J_3=1.95$.
The inset enlarges the scale to make the phase transitions ${\rm I}\to {\rm II}\to {\rm III}$ more clearly visible.
 }
\label{fig_BM195}
\end{figure}

There are four phases ${\rm I}$ to ${\rm IV}$, such that
${\rm I},{\rm II}$ and ${\rm III}$ are non-coplanar and $IV$ is coplanar, see Figure \ref{fig_PD2}.
Accordingly, there are two types of magnetization curves,
the first one corresponding to the sequence of phase transitions
${\rm I}\stackrel{{\mathcal J}}{\to}{\rm II}\stackrel{{\mathcal K}}{\to}{\rm III}\stackrel{{\mathcal K}}{\to}{\rm IV}\stackrel{{\mathcal K}}{\to}FM$,
for $J_3>1.52$, see Fig.~\ref{fig_BM195} for the example $J_3=1.95$.
The second type of  magnetization curves corresponds to the sequence of phase transitions
${\rm I}\stackrel{{\mathcal J}}{\to}{\rm III}\stackrel{{\mathcal K}}{\to}{\rm IV}\stackrel{{\mathcal K}}{\to}FM$,
for  $J_3<1.52$.

The four phases of the FM/AF square kagom\'{e} show certain differences to the AF case.
Phase ${\rm I}$ again consists of 14 spins, but this time arranged in three squares and a pair,
so that for $H\to 0$ one square degenerates into a pair.
Phase ${\rm III}$ has the same structure but different energy.
Phase ${\rm II}$, which squeezes between phase $I$ and phase ${\rm III}$ for $J_3>1.52$
consists of $7$ pairs. At the transition ${\rm III}\stackrel{\mathcal K}{\to} {\rm IV}$,
one rectangle and one pair converge towards the north pole,
while the other two rectangles merge into one pair.
This constitutes the analytical phase ${\rm IV}$ with three coplanar spins,
all of which converge towards the north pole for $H\to H_{\rm sat}$.

\section{Conclusions}

We have analyzed classical spin-lattice models with cuboctahedral ground states,
and compared magnetization curves and phase transitions in different systems.
The focus here is on the kagom\'{e} models,
where the classical magnetization curves have not yet been investigated for the $J_1 - J_2$ variant.
In addition, we have compared our findings on the $J_1-J_d$ kagom\'{e} model
with published results and recapitulated our recent study on the
$J_+=J_\times$ square kagom\'{e} lattice. Typical phenomena are different types
of deformed cuboc ground states for moderate magnetic fields and, in most cases,
double umbrella states approaching the fully aligned state at high fields.
Key features of the magnetization process can also be studied using a
simplified 12-spin model, which we propose in this work.
Interestingly, in the 12-spin model, there are regions of the magnetic field
where groups of spins rotate in the opposite direction to the field direction as $H$ is increased.

The methods used include numerical ground state determination by iterative minimization,
a semi-analytical approach and in some cases, especially for the last phase
before saturation, fully analytical solutions.
These methods allow us to obtain very accurate phase boundaries and phase diagrams.

In summary, our research sheds light on the magnetization process of classical
Heisenberg magnets with non-coplanar cuboc ground states and reveal a complicated
behavior influenced by the lattice geometry and exchange couplings.
This study will hopefully not only improve our understanding of frustrated magnetism,
but also contribute to the broader study of exotic magnetic phenomena.\\

\appendix

\section{Construction of the 12 spin model}\label{sec:SSM12}
We recall the problem of defining coupling constants $J_{i,j}$ between $N=12$ spins
such that the resulting ground state forms a cuboctahedron in spin space.
The vertices ${\mathbf c}_i,\;i=1,\ldots ,12$ of the cuboctahedron can be chosen
as those 12 vectors with exactly one zero component and the other two components
being $\pm 1$. The normalization factor is irrelevant for the moment. The numbering
of the vertices is arbitrary, but for the Fig.~\ref{fig_model_i_TOY}, left panel,
a certain choice of the numbering has been made.
We represent these 12 vertices as rows of a $12\times 3$ - matrix $C$ with
entries $C_{ik},\,i=1,\ldots,12,\;k=1,2,3$.
Then we define a projector $P$ in the 12-dimensional space ${\mathbb R}^{12}$ by
\begin{equation}\label{defP}
  P_{ij}=\frac{1}{8}\sum_{k=1}^{3} C_{ik}\,C_{jk}
  \;.
\end{equation}
It projects onto the  $3$-dimensional subspace of ${\mathbb R}^{12}$
spanned by the $3$ columns of $C$.
The symmetric matrix of coupling constants
\begin{equation}\label{defJ}
 J:= 2\, {\mathbbm 1}- 8\,P
\end{equation}
has zero trace and, moreover, zero diagonal entries.
Its non-zero entries are of the form $J_1=-1$, $J_2=1$, $J_3=2$,
as announced in Section  \ref{sec-3}.
By construction, the eigenvalues of $J$ are $-6$ ($3$-fold degenerate)
and $2$ ($9$-fold degenerate). The ground state of the 12 spin model
is hence the unique $3$-dimensional cuboc state that can be obtained by linear combinations
of the eigenvectors corresponding to the lowest eigenvalue $-6$ of $J$.

This method to construct spin systems with given ground states can be generally
applied and is not confined to the 12 spin system under consideration.

%


\end{document}